\documentclass{kluwer}    

\newdisplay{guess}{Conjecture}
\usepackage[dvips]{epsfig}

\begin{document}
\begin{article}
\begin{opening}
\title{UV Upturn in Elliptical Galaxies: Theory}
\author{Sukyoung K. \surname{Yi} and Suk-Jin Yoon}
\runningauthor{Sukyoung K. Yi} \runningtitle{UV Upturn in
Early-Type Galaxies and EHB Stars} \institute{University of
Oxford, Astrophysics}
\date{Published in 2004,Astrophysics and Space Sciences, 291, 205}

\begin{abstract}
I review the current theoretical understanding on the UV upturn
phenomenon in early-type galaxies.
\end{abstract}

\end{opening}

\section{History}

The UV upturn is the rising flux with decreasing wavelength
between the Lyman limit and 2500\AA\, found virtually in all
bright spheroidal galaxies. It has been a mystery ever since it
was first detected by the OAO-2 space telescope (Code \& Welch
1979) because such old metal-rich populations were not expected to
contain any substantial number of hot stars. It was confirmed by
following space missions, ANS (de Boer 1982), IUE (Bertola et al.
1982) and HUT (Brown et al. 1997). The positive correlation
between the UV-to-optical colour (i.e., the strength of the UV
upturn) and the Mg2 line strength found by Burstein et al. (1987)
through IUE observations has urged theorists to construct novel
scenarios in which metal-rich ($\gtrsim Z_{\odot}$) old ($\gtrsim$
a few Gyr) stars become UV bright (Greggio \& Renzini 1990; Horch
et al. 1992). Also interesting was to find using HUT that,
regardless of the UV strength, the UV spectral slopes at
1000--2000\AA\, in the six UV bright galaxies were nearly
identical suggesting a very small range of temperatures of the UV
sources in these galaxies (Brown et al. 1997), which corresponds
to $T_{\rm eff} \approx 20,000 \pm 3,000$\,K. This, together with
other evidence, effectively ruled out young stars as the main
driver of the UV upturn. A good review on the observational side
of the story is given in the next article by Tom Brown, as well as
in the recent articles of Greggio \& Renzini (1999) and O'Connell
(1999).

\section{Theory}

Theorists aim to present a model that explains three basic
observational facts: (1) UV upturn being ubiquitous in the cores
of bright elliptical galaxies whose $optical$ stellar lights
appear to be dominated by old and metal-rich stars, (2) positive
correlation between the strength of the UV upturn and the
$optical$ metal line (Mg2) strength, and (3) a narrow range of
temperature of UV sources.

\subsection{Metal-poor HB hypothesis}

It is widely known that metal-poor HB stars can be hot and make
good UV sources when they are old (see the article by Yoon in this
volume). Thus, the first scenario was naturally that bright
elliptical galxies contain a fraction (perhaps order of 20\%) of
stellar mass in the core in extremely old and metal-poor
populations (Park \& Lee 1997). The strength of this scenario is
that the oldest stars in a galaxy must also be the most metal-poor
and are likely to be in the core, where the UV upturn is found.
Besides, such old metal-poor HB stars are nothing too exotic to
believe. In this scenario, the UV vs Mg2 relation does not present
causality but simply a result of tracing different populations in
terms of metallicity. The narrow range of temperature is very well
reproduced as well. On the other hand, the mass fraction of order
$~$20\% is too high to be allowed by the standard galactic
chemical evolution theory. If they are present at that level,
there must also be a large number of intermediate-metallicity
(20--50\% solar), which will make galaxy core's integrated
metallicity too low and its integrated colours too blue, compared
to observed values. Moreover, the age of the oldest stars, i.e.
the main UV sources, must be 30\% older than the average Milky Way
globular clusters. Cosmology and stellar evolution theory (mainly,
on globular clusters) have just reached a common ground where they
approve each other in terms of the age of the universe. But now
this scenario requires a large cosmological constant
$\Omega_\Lambda \gtrsim 0.8$ and poses a problem to the apparent
peace. This called for alternative scenarios.

\subsection{Metal-rich HB hypothesis}

Stellar evolutionists found that metal-rich HB stars could in fact
be more effective UV bright sources if galactic helium is enriched
with respect to heavy elements at a rate of $\Delta Y$/$\Delta Z
\gtrsim 2.5$ or if the mass loss rate in metal-rich stars is
30--40\% higher than that of metal-poor stars (Horch et al. 1992;
Dorman et al. 1995; Yi et al. 1997a). Both of these conditions are
difficult to validate empirically but quite plausible (Yi et al.
1998). In this scenario, metal-rich stars become UV bright in two
steps: (1) they lose more mass on the red giant phase due to the
opacity effect and become low-mass HB stars, and (2) extremely
low-mass HB stars stay in the hot phase for a long time and
directly become white dwarfs, effectively skipping the red,
asymptotic giant phase (Yi et al. 1997a, 1997b). This scenario
reproduces most of the features of the UV upturn (Bressan et al.
1994; Yi et al. 1998). The UV vs Mg2 relation is also explained.
On the other hand, its validity heavily hinges upon the
purely-theoretical late-stage stellar evolution models of
metal-rich stars.

\subsection{Which theory?}

Both of these scenarios are equally appealing but their
implications on the age of E galaxies are substantially different.
The metal-poor hypothesis suggest UV- strong galaxies are 30\%
older than the Milky Way (MW) and requires the universe to be
older than currently believed, suggesting a large cosmological
constant. The metal-rich hypothesis on the other hand suggests
that E galaxies are not necessarily older than the Milky Way halo.

\section{Obsrvational tests}

The first obvious question is whether we can measure the mean
metallicity of the UV sources from  the high resolution UV spectra
of the UV-upturn galaxies. Brown et al. (2002) derived roughly
10\% solar from the HST/STIS data on NGC\,1399 but found the
derived value misleading because such highly-evolved hot HB stars
do not show their intrinsic metallicities due to various heavy
element redistribution processes. Spectroscopy does not seem
promising at the moment.

Yi et al. (1999) presented a redshift-evolution test. Though both
the metal-poor and metal-rich hypotheses are calibrated to the
$z=0$ galaxies, they predict different evolutionary paths in the
UV strength as a function of time. Thus, by reaching as far as $z
\approx 0.3$, one might be able to select the more likely model.
Figure 1(a) shows their $qualitative$ prediction. Brown et al.
(1998, 2000, 2003) obtained UV data on three galaxy clusters at
$z$ = 0.33, 0.375, 0.55. Figure 1(b) shows them compared to our
metal-rich HB hypothesis models. The models are denoted by the
shaded region defined by the spread in the metallicity by a factor
of two (the more metal-rich models, the more UV-strong). The
earliest data on the $z=0.375$ cluster (shown as triangles) appear
too red to be fit by the models and show no significant evolution
in the UV flux from $z=0$, which surprised many theorists
(including me). The more recent two cluster data with proper error
estimates however look consistent with the metal-rich HB
hypothesis models. Brown at this meeting cautiously stated that
the first data set at $z=0.375$ obtained using HST/FOC may be
dubious. It is encouraging to see the models mimic the two cluster
data reasonably well, but we definitely need more data.

\begin{figure}
\tabcapfont
\centerline{%
\begin{tabular}{c@{\hspace{0.pc}}c}
\includegraphics[width=2.4in]{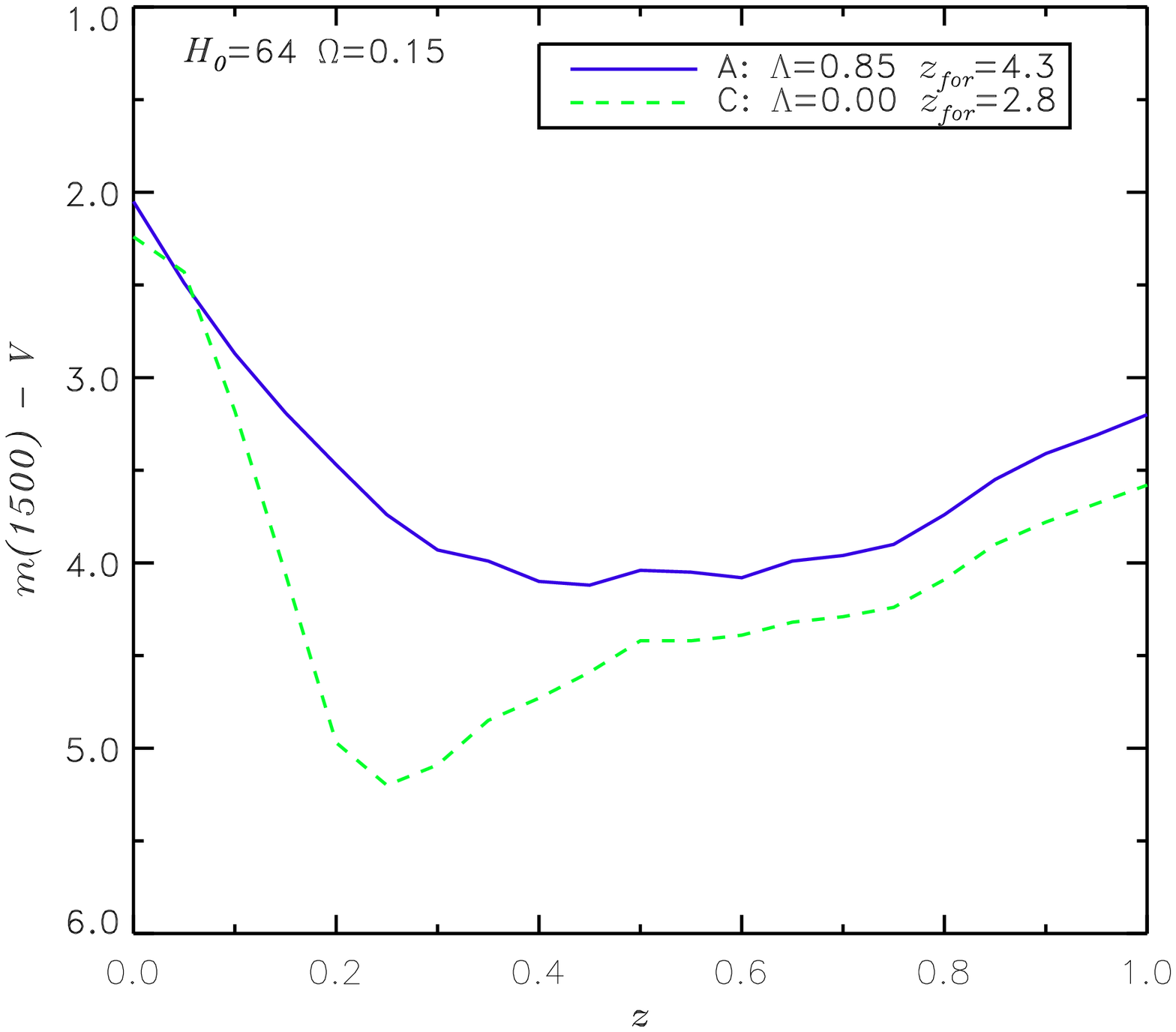} &
\includegraphics[width=2.4in]{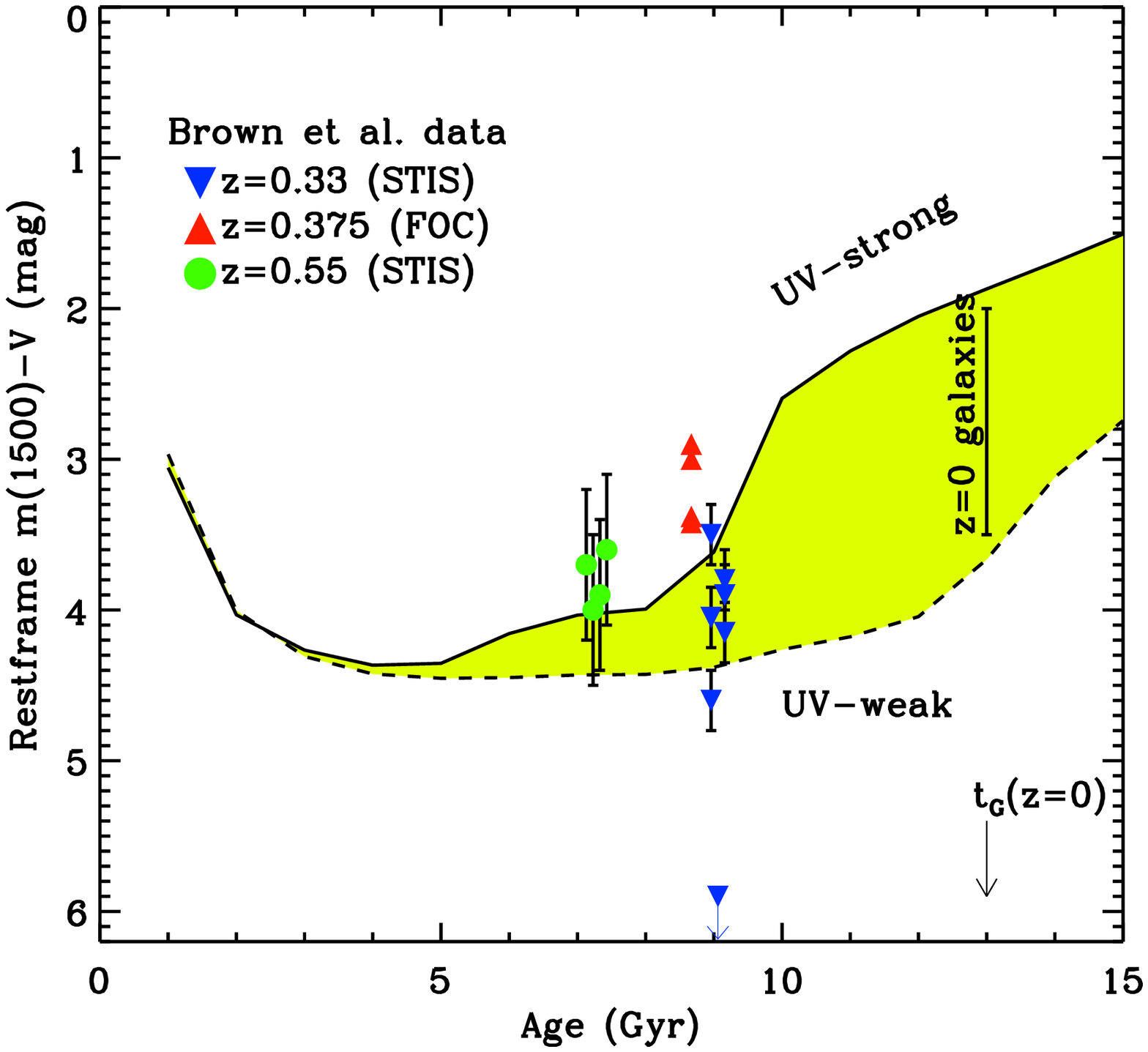} \\
a. predicted $z$-evolution  & b. comparison with data
\end{tabular}}
\caption{ (a) The two representative UV evolution models from Yi
et al. (1999). Model A: metal-poor HB (older) hypothesis, Model C:
metal-rich HB (younger) hypothesis. (b) The HST data (triangles
are now considered unreliable) seem consistent with the models
(here shown are metal-rich models), but more data are required.
}\label{uvupturn}
\end{figure}

\section{Issues}

Readers may get an impression by reading the previous sections
that we have solid and successful theories. Quite contrarily,
there are several critical issues to be understood before we can
ever claim so.

\subsection{$\alpha$-enhancement}

Theorists (including myself) often interpret the UV vs Mg2
relation as a metallicity effect on the UV flux. However, it
should be noted that Mg2 strength may not be representative of the
overall metallicity. In fact, it has been known that elliptical
galaxies are enhanced in $\alpha$-elements with respect to iron.
We then naturally wonder if it is not the overall metallicity but
$\alpha$-enhancement that generates the UV upturn. To perform this
test, we need $\alpha$-enhanced stellar models. The $Y^2$
Isochrones group have released their stellar models for the main
sequence (MS) through red giant branch (RGB) (Yi et al. 2003, see
also my another article in this volume). No $\alpha$-enhanced HB
models are available yet. $\alpha$-enhancement can have several
impacts on the galaxy spectral evolution. First, it changes the
stellar evolutionary time scale, as CNO abundance affects the
nuclear generation rates. Second, it changes opacities and thus
the surface temperatures of stars. These two effects will make a
change in the mass loss computed using a parameterised formula,
such as the Reimers formula. For a fixed mass loss efficiency, we
find the $\alpha$-enhanced ([$\alpha$/Fe]=0.3--0.6) tracks yield
$\approx 0.03 M_{\odot}$ smaller mass loss at ages 5--8Gyr but
$\approx 0.03 M_{\odot}$ greater mass loss at ages
$\gtrsim8$\,Gyr, compared to the standard ([$\alpha$/Fe]=0)
tracks. $\alpha$-enhancement must have similar opacity effects on
the HB evolution except that mass loss on the HB is negligible.
Thus its effects are expected to be greater to the MS to RGB than
to the HB phase. Considering all these, I decided to inspect the
overall effects of $\alpha$-enhancement by just adopting new
$\alpha$-enhanced MS through RGB tracks, ignoring the change in
the HB models. Figure 2 shows the results for two metallicities
and three values of $\alpha$-enhancement. In old metal-poor model
(Fig.~2(a)) $\alpha$-enhancement causes a positive effect to the
$relative$ UV strength because (1) it causes a slight increase in
mass loss on the RGB and (2) it causes MS stars and red giants to
be redder and fainter in $V$ band. The [$\alpha$/Fe]=0.3 model
roughly reproduces the SED of a typical UV-strong metal-poor
globular cluster, which is quite satisfying. The metal-rich models
(b) do not show any appreciable change in response to
$\alpha$-enhancement. Because giant E galaxies are largely
metal-rich (roughly solar) and the light contribution from
metal-poor stars even in the core is not substantial, this
exercise by itself does not seem to support the idea that
$\alpha$-enhancement is the main driver of the UV upturn.

\begin{figure}
\tabcapfont
\centerline{%
\begin{tabular}{c@{\hspace{0.pc}}c}
\includegraphics[width=2.4in]{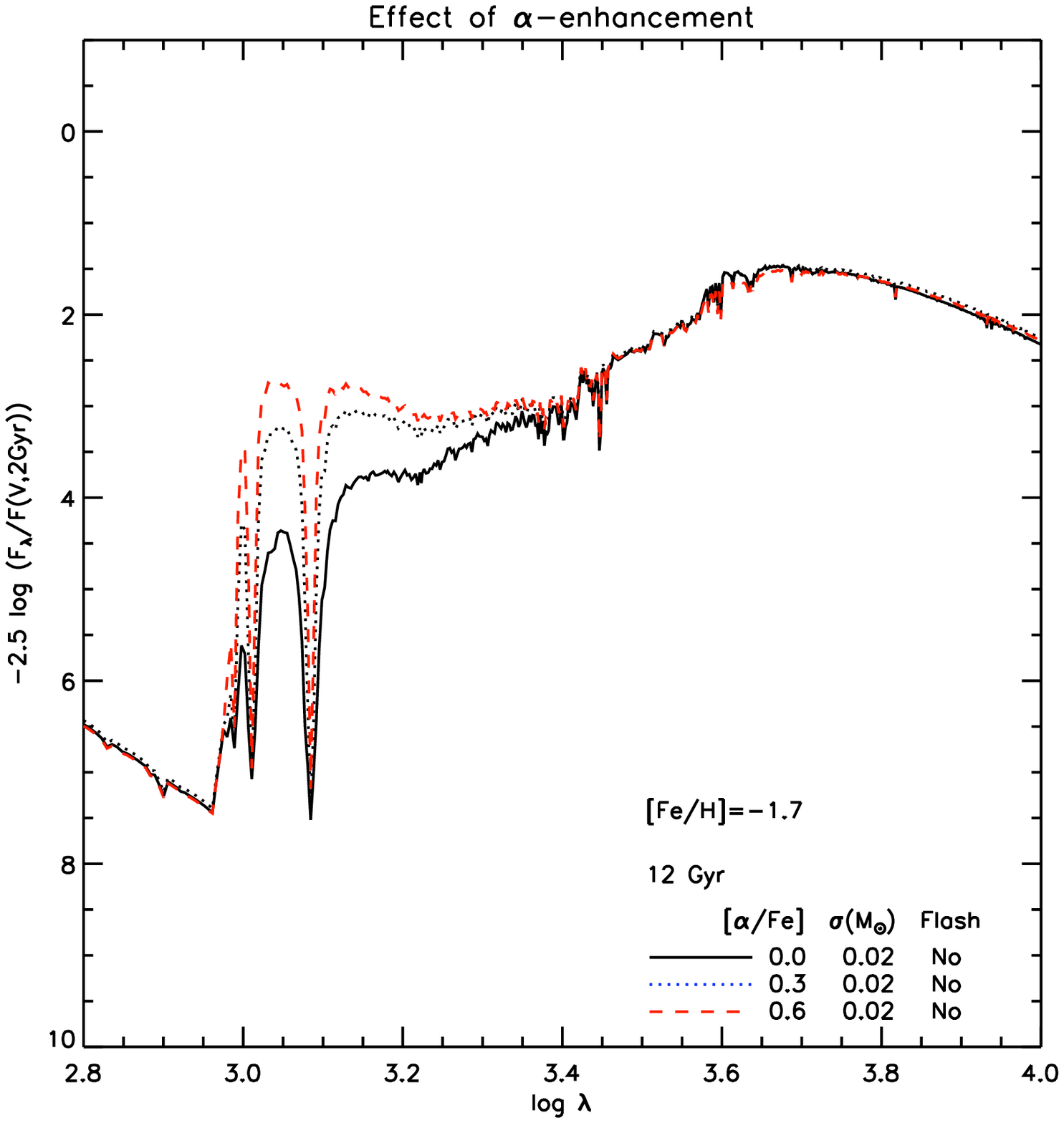} &
\includegraphics[width=2.4in]{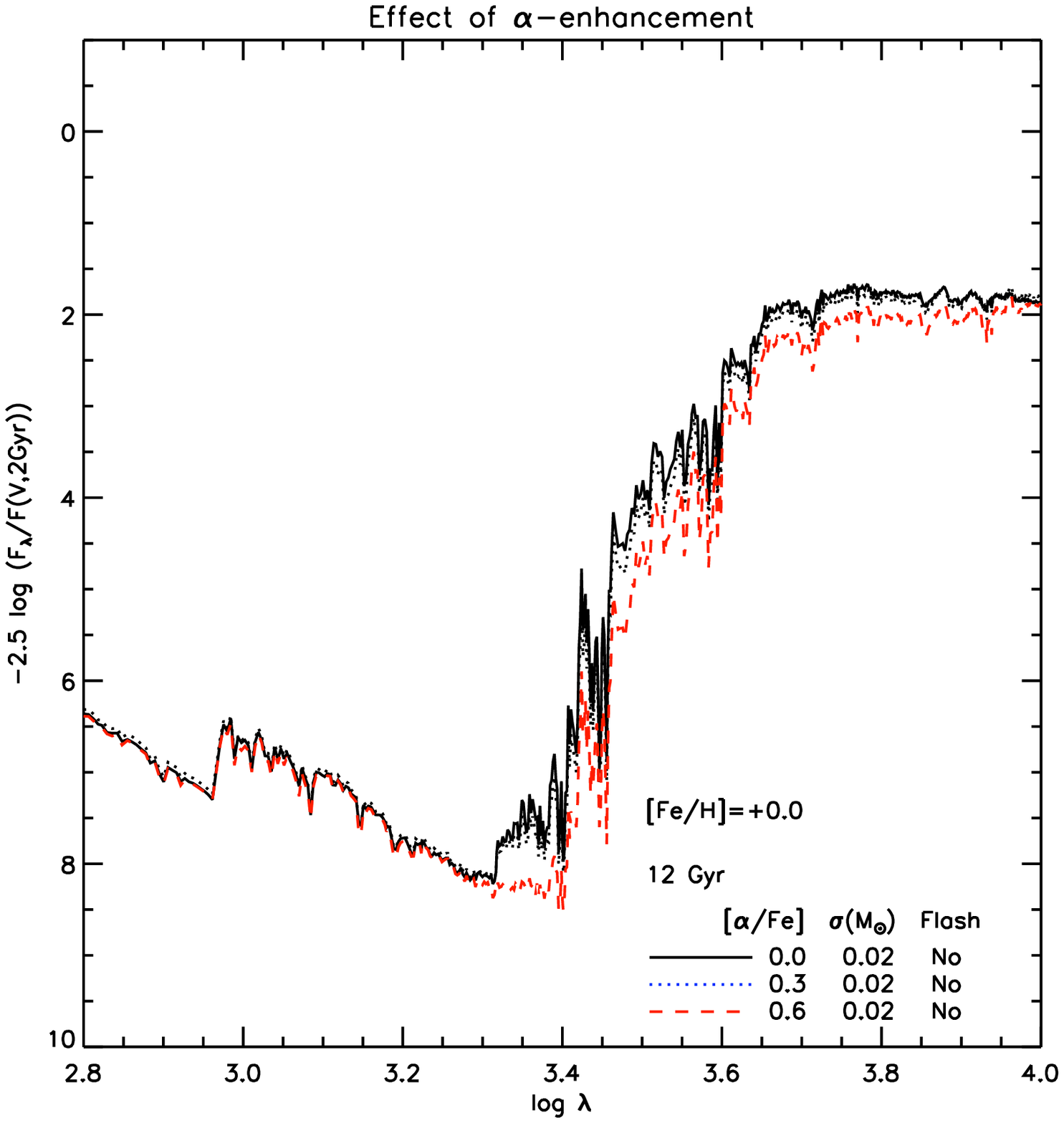} \\
a. predicted $z$-evolution  & b. comparison with data
\end{tabular}}
\caption{ (a) The two representative models from Yi et al. (1999).
Model A: metal-poor HB (older) hypothesis, Model C: metal-rich HB
(younger) hypothesis. The two models predict different
$z$-evolution. (b) The HST data (triangles are now considered
unreliable) seem consistent with the metal-rich models, but more
data are required. }\label{ae}
\end{figure}

\subsection{EHB stars in globular clusters}

With the HST spatial resolution, a number of studies have found
hot, extended horizontal branch (EHB) stars in globular clusters
(e.g., Piotto et al. 1999). They are efficient UV sources and
could be important candidates as the main UV sources in E
galaxies; but there is no population synthesis model that
successfully reproduces them as they are observed (number density,
colours and brightness).

\subsection{NGC 6791}

This old (8-9Gyr) metal-rich (twice solar) cluster is unique for
being close to the stellar populations of the E galaxy cores.
Strikingly, roughly 9 out of its 32 HB stars have the properties
of typical EHB stars (Kaluzny \& Udalski 1992; Liebert et al.
1994). Figure 3(a) shows the well-defined red clump (compared to a
synthetic HB shown as diamonds) and several hot (probably EHB)
stars shown in box (see Yong et al. 2000 for details). Note that
the synthetic HB does not contain any hot HB stars, that is, no
diamond away from the clump. It is critical to understand the
origin of these {\em hot old metal-rich stars}. Landsman et al.
(1998), based on UIT data, concluded that NGC 6791, if observed
from afar without fore/background stellar contamination, would
exhibit a UV upturn just like the ones seen in E galaxies.

Through detailed synthetic HB modelling I have found that it is
impossible to generate an HB with such a severely-bimodal colour
distribution as shown in this cluster, unless an extremely (and
unrealistically) large mass dispersion is adopted. In the hope of
finding a mechanism that produces such an HB Yong et al. (2000)
explored the effect of mass loss $on$ the HB. Yong et al. found
that with some mass loss taking place on the HB ($\approx
10^{-9}-10^{-10}\,M_{\odot}\,yr^{-1}$) HB stars born cool quickly
become hot, suggesting that mass loss on the HB might be an
effective mechanism of producing such stars. Figure 3(b) shows a
sample result. The top panel shows the observed colour
distribution, red clump stars being simply put as an asterisk. The
middle panel shows a synthetic HB assuming
$10^{-9}\,M_{\odot}\,yr^{-1}$, with a Reimers mass loss efficiency
$\eta = 1$ ($\approx$ 40\% greater than that derived from
metal-poor globular cluster stars) and a gaussian mass dispersion
parameter $\sigma = 0.04\,M_{\odot}$. The bottom panel compares
the distributions from theory and observation; the agreement is
quite good. Do we have a convincing theory? If so, what causes
this mass loss?

Vink \& Cassisi (2002) however pointed out that the scale of the
mass loss assumed by Yong et al. is not justified by their
radiation pressure calculations. It seems that such mass loss is
not supported in single stellar evolution. Green et al. (2000)
reported that most of these hot stars in NGC 6791 are in binary
systems. If they are close binaries and experience mass transfer
it would be an effective mechanism for mass loss on the HB as well
as on the RGB.

\begin{figure}
\tabcapfont
\centerline{%
\begin{tabular}{c@{\hspace{0.pc}}c}
\includegraphics[width=2.in]{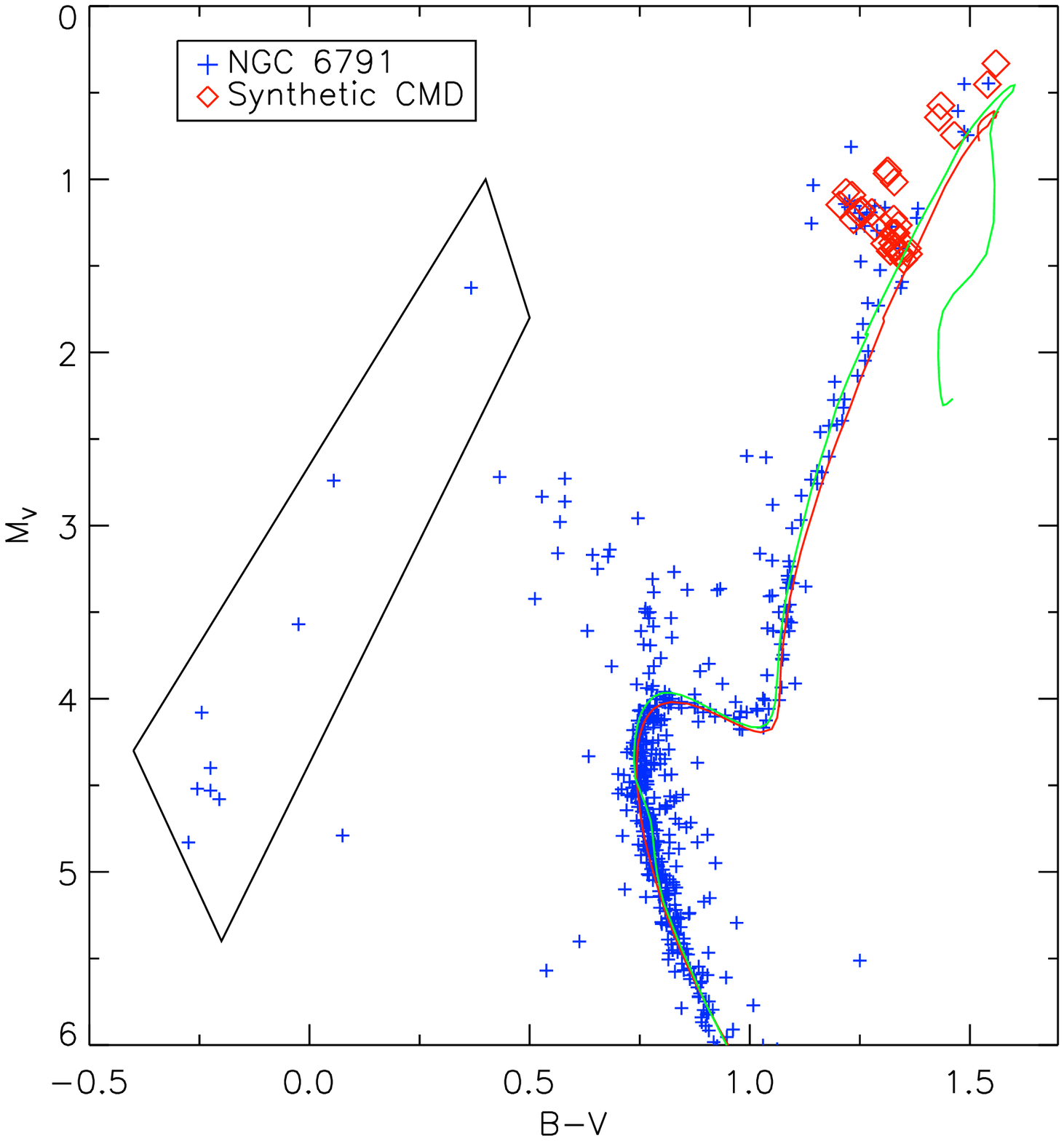} &
\includegraphics[width=2.6in]{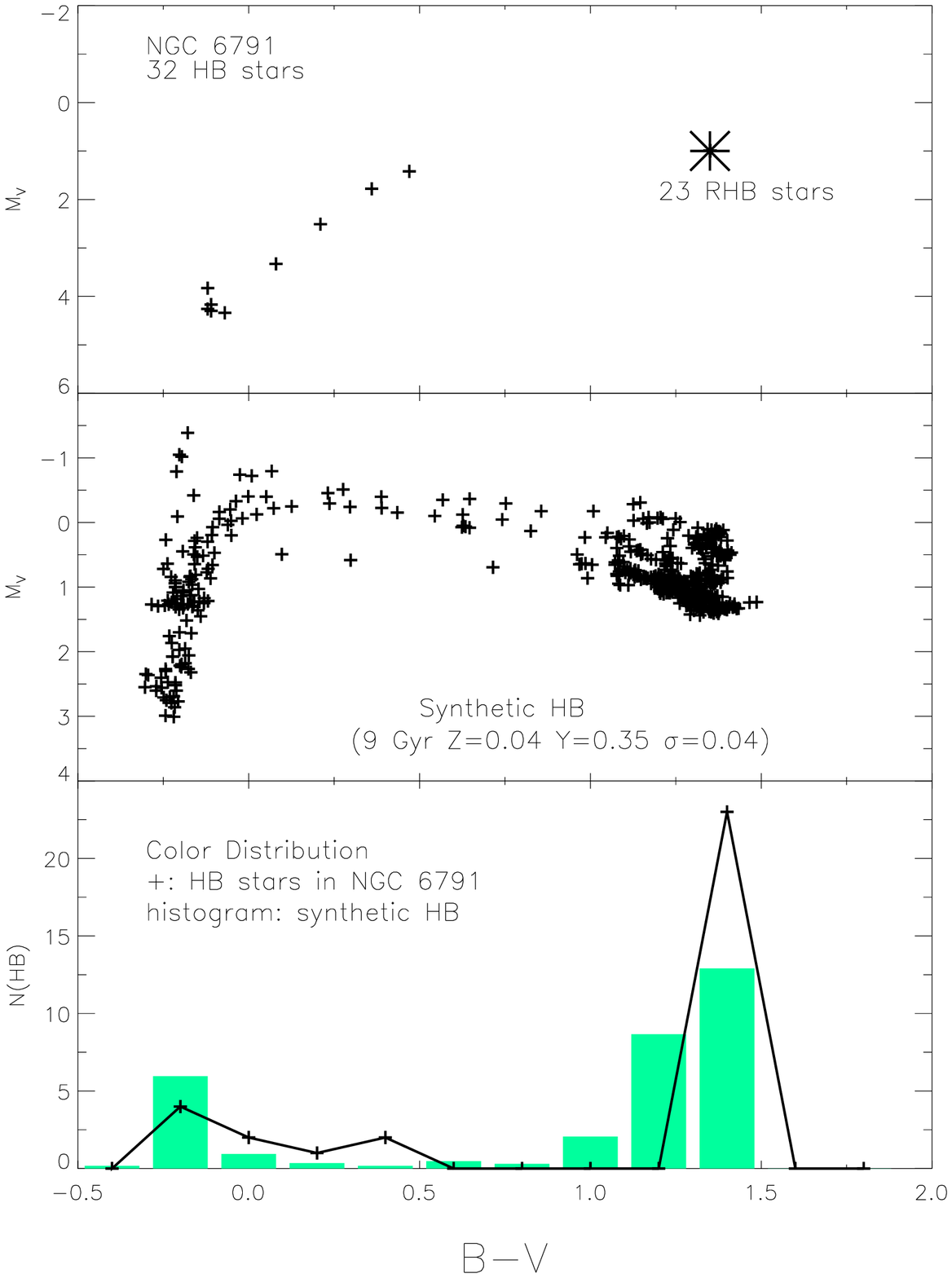} \\
a. NGC 6791  & b. Colour distribution
\end{tabular}}
\caption{ (a) NGC 6791. The hot stars (shown in box) are not
reproduced by canonical models (diamonds). (b) With mass loss
assumed to take place on the HB the colour bimodality is
reproduce.} \label{n6791}
\end{figure}

\subsection{Binaries}

SdB stars may be the field counterparts of the EHB stars in
clusters. They also have the properties that are similar to those
of the UV sources in the UV-upturn galaxies. More than 70\% of sdB
stars are found to be in binary systems (Saffer et al. 2000;
Maxted et al. 2001), and we may be looking at a smoking gun.

Han et al. (2003) used binary population synthesis technique to
study the effects of binary evolution and found that 75--90\% of
sdB stars should be in binaries. SdBs are detected to be in a
small mass range centered at 0.5\,$M_{\odot}$, but Han et al.
found that the range should be in truth as wide as 0.3 through 0.8
$M_{\odot}$. They predict a birthrate of 0.05~$yr^{-1}$ for
Population I stars and 6 million sdB stars in the disc. Assuming
the Galactic Disc mass of 5x$10^{10}$\,$M_{\odot}$, this means
roughly 100 sdB stars per $10^6 M_{\odot}$. In a
back-of-the-envelope calculation, there are roughly a few thousand
HB stars per million solar mass in globular cluster populations.
The comparison between the sdB rate (100 per $10^6 M_{\odot}$) and
that of the HB (say, 5000 per $10^6 M_{\odot}$) suggests an old
disc population may develop 1 sdB star for 50 HB stars (2\%). This
is hardly impressive. NGC 6791 shows roughly 30\% and the most
UV-bright globular cluster $\omega$\,Cen 20\%. Yet, even
$\omega$\,Cen does not exhibit a UV upturn. If this calculation is
realistic at least within an order, binary mass transfer may not
provide the origin of the majority of the UV sources in UV-upturn
galaxies. On the other hand, a larger sdB production rate might be
plausible in elliptical galaxy environment due to large age and/or
large metallicity. A more detailed study is very tempting.

\subsection{Other issues}

There are other important issues as well.  For example, the
late-stage flash mixing scenarios and the like (D'Cruz et al.
1996; Brown et al. 2001) may also be effective ways of producing
hot stars (such as sdB stars) in old populations. Their typical
temperature range and the predicted birthrate may not be entirely
consistent with the UV upturn shown in E galaxies, however.

Another important observational constraint comes from the HST UV
images of M32. First, Brown et al. (2000) found that PAGB stars
are two orders of magnitudes fewer than predicted by simple
stellar evolution theory. This is significant as PAGB stars may
contribute 10--30\% of the UV flux in the UV-upturn galaxies. More
importantly, they find too many hot HB stars to reproduce with
standard population models that are based on the mass loss rate
calibrated to the globular cluster HB morphology ($\eta \approx
0.65$). It is possible to reproduce the observed number of hot
stars in M32-type populations if a greater mass loss rate is used,
which would be consistent with the variable mass loss hypothesis
of Yi et al. (1997b, 1998).

\section{Summary}

The UV upturn provides tests of various aspects of stellar
evolution and constrains the ages of elliptical galaxies. The
current population synthesis models reasonably reproduce UV to
optical broadband colours of current epoch galaxies, but there are
multiple solutions. Redshift evolution observations have been
proposed to differentiate these models. The HST/STIS instrument is
being used effectively (Brown and collaborators) and the GALEX
(Martin et al. 1998) has a program to explore this subject.

As soon as we look at the observed UV continua (spectra rather
than colours) and the models, we realise something is not quite
right with the models yet. The detailed spectral shape comparison
clearly shows that there should be a stronger temperature
bimodality in the HB than current models suggest. This seems
related with the colour bimodality in the HB and the presence of
the EHB stars in globular clusters found in many star clusters.
The origin of sdB stars may also provide a clue. GALEX All Sky
Survey will detect nearly all sdB stars within 25kpc at Galactic
latitude of roughly 30 or greater. The data may pin down the
number density and birthrate of sdB stars further. A more fine
tuning on the mass loss calibration, as evident from M32
observations, seems required as well. Regarding the composite
nature of galaxies, the colour bimodality may also very well arise
from a large age difference between metal-poor (older) and
metal-rich (younger) populations. When we have a better
understanding on these issues, we will be able to find a
convincing theory at last.

\acknowledgements I thank the Keele Conference Local Organising
Committee for the exciting meeting (with ample discussion) and for
inviting me to be part of the Science Organising Committee and
deliver this talk.


\theendnotes

\end{article}
\end{document}